# To Block or Not to Block: Accelerating Mobile Web Pages On-The-Fly Through JavaScript Classification

Moumena Chaqfeh
NYUAD
moumena@nyu.edu

Muhammad Haseeb
LUMS
haseebluminite@gmail.com

Waleed Hashmi
NYUAD
waleedhashmi@nyu.edu

Patrick Inshuti
NYUAD
pim219@nyu.edu

Manesha Ramesh
NYUAD
mr4684@nyu.edu

Matteo Varvello
Nokia Bell Labs
varvello@gmail.com

Fareed Zaffar
LUMS
fareedz@gmail.com

Lakshmi Subramanian
NYU
lakshmi@cs.nyu.edu

Yasir Zaki
NYUAD
yasir.zaki@nyu.edu

## ABSTRACT

The increasing complexity of JavaScript in modern mobile web pages has become a critical performance bottleneck for low-end mobile phone users, especially in developing regions. In this paper, we propose SlimWeb, a novel approach that automatically derives lightweight versions of mobile web pages on-the-fly by eliminating the use of unnecessary JavaScript. SlimWeb consists of a JavaScript classification service powered by a supervised Machine Learning (ML) model that provides insights into each JavaScript element embedded in a web page. SlimWeb aims to improve the web browsing experience by predicting the class of each element, such that essential elements are preserved and non-essential elements are blocked by the browsers using the service. We motivate the core design of SlimWeb using a user preference survey of 306 users and perform a detailed evaluation of SlimWeb across 500 popular web pages in a developing region on real 3G and 4G cellular networks, along with a user experience study with 20 real-world users and a usage willingness survey of 588 users. Evaluation results show that SlimWeb achieves a 50% reduction in the page load time compared to the original pages, and more than 30% reduction compared to competing solutions, while achieving high similarity scores to the original pages measured via a qualitative evaluation study of 62 users. SlimWeb improves the overall user experience by more than 60% compared to the original pages, while maintaining 90%-100% of the visual and functional components of most pages. Finally, the SlimWeb classifier achieves a median accuracy of 90% in predicting the JavaScript category.

## 1 INTRODUCTION

JavaScript (JS) has proven to be the most expensive resource for web browsers to process and a critical factor in performance degradation of web pages [64]. The impact of JS is even worse for a large fraction of users around the globe who live in low- and middle-income countries, where mobile accounts for 87% of the total broadband connections [13], and users solely rely on affordable low-end smartphone devices to access the web [3]. On these devices, JS processing time is tripled compared to desktops [36], resulting in long delays [64] and a poor browsing experience [57, 70]. In contrast with other Internet applications (such as video streaming), the performance of web browsing is more sensitive to low-end hardware [34]. Nevertheless, the current status of the World Wide Web (WWW) shows a 45% increase in JS usage by a median mobile page, with only 7% less JS kilobytes transferred to mobile pages in comparison to desktop pages (475.1 KB for desktop and 439.9 KB for mobile) [15].

Surprisingly, although the main objective of JS in web pages is to provide interactive contents, and despite the growing usage of JS in web pages, JS is not always essential to the main page content or interactive functionality [27], and pages can be served with the same original structure and full interactivity while excluding many of JS elements. From the user's point of view, if the main contents of the pages are preserved and the functional elements are responding properly, then why should a user with low-end settings tolerate the extra loading time due to a set of analytic elements (for example)? From a page developer's perspective, it would be better to satisfy the requirements of more users by offering lighter versions of web pages, instead of losing them due to computationally intensive pages [31].

This paper proposes SlimWeb, a JS classification approach that lightens mobile web pages *on-the-fly* with the aid of a browser's plugin. This is achieved by preserving critical JS elements on these pages, while blocking the non-critical ones. What makes SlimWeb unique, in comparison to state-of-the-art solutions, is its ability to determine the "criticality" of JS elements without the need to execute their code. Additionally, SlimWeb maintains the core visual, interactive and functional components of the mobile web pages.

SlimWeb is realized as a service that periodically crawls popular web pages and classifies their embedded JS – using a supervised learning model – into *critical* and *non-critical* elements. SlimWeb shares the classification results with end-users though the browser plugin. The plugin is responsible for preserving the critical JS elements while blocking the non-critical ones, effectively producing lighter mobile pages. SlimWeb is primarily designed for low-end smartphones which are common among mobile users in developing regions [8]. SlimWeb is built on the observation that reducing JavaScript bundles on a web page is the best way to improve its performance [36]. Beyond improving Page Load Times (PLT), reducing non-critical JavaScript can also reduce bandwidth and energy consumption on mobile devices, especially given that web browsing is known to cause major drain on the battery lifetime [57].

JavaScript classification is a challenging problem due to the dynamic non-deterministic behavior of the code [67], especially when attempted on-the-fly using low-end mobile phones, since this leaves no room for executing the JS code– which tends to consume numerous resources and time. Existing research on JS filtering primarily focuses on the identification of advertising and/or tracking JS [42, 47, 49], in addition to malicious JS code snippets [32, 33, 44, 66, 68, 69]. Existing client-side solutions to reduce the cost of JS rely on the integration of a JS blocker as a browser extension [35, 58]. However, these blockers aim to prohibit a specific category of JS (such as Ads) [6, 18, 24], by relying on predefined blocking lists. Although these lists can be periodically updated, existing blockers fail to automate the classification of previously-unseen JS elements that can be created from existing known libraries by changing the serving domain or obfuscating code and metadata of web pages [17].

With SlimWeb, we aim to fill this gap by classifying JS elements via a supervised Machine Learning (ML) model. Instead of examining the JS code to infer its semantics and behaviour, SlimWeb extracts highly predictive features from the code, and then uses them to classify JS. A labeled dataset is required to train such a JS classifier, which unfortunately does not exist today to the best of our knowledge. Thus, using categories identified by experts of the web community [22], we built a labeled dataset with 127,000 JS elements by scraping more than 20,000 popular web pages. Using this labeled dataset, we show that SlimWeb's JS classifier achieves a 90% classification accuracy across eight different JS categories using 508 carefully selected features.

To answer the challenging question of which categories are critical and which are not, we surveyed 306 users from different countries to set SlimWeb plugin defaults. Results show that the majority of users agree that *advertising* and *analytics* are non-critical regardless of their network connectivity or mobile phone class. Utilizing different settings, we performed a quantitative and a qualitative evaluation of SlimWeb. The quantitative evaluation was conducted by installing the SlimWeb plugin in a low-end mobile phone browser, from which 500 popular mobile pages were requested through real 3G and 4G cellular networks. The qualitative evaluation was conducted by collecting scores given by 40 users to evaluate 95 popular pages. Evaluation results show a 50% reduction in the page load time and an improvement in the overall user experience of more than 60%, while maintaining more than 90% of the visual contents and interactive functionality of most pages. We conclude our evaluations with a user experience study conducted by 20 users over a period of two weeks. All participants had consensus that SlimWeb provided faster and lighter pages. Additionally, they gave a median score of 80% in rating the content completeness, as well as retaining the functional elements of the pages, without raising issues related to broken pages. In case of a broken page, SlimWeb plugin allows to disable blocking on a per-page basis. We asked 588 users from different countries about their willingness to put in the effort required for fixing these pages, and the majority of them were willing to do so. In summary, the contributions of this paper include:

- An ML-driven JS classifier that categorizes JS elements with 90% accuracy.
- An easy to install plugin that benefits from the classification in providing lighter versions of web pages for mobile users.
- 50% reduction in page load times compared to the original pages, as well as more than 30% reduction in comparison to state-of-the-art solutions.
- More than 60% improvement in the user experience, while maintaining more than 90% similarity to the original pages.

## 2 MOTIVATION

In this section, we motivate the core design principle of SlimWeb, which identifies and removes non-critical JS in web pages to significantly improve the user experience without impacting the functionality of the pages.

### 2.1 Non-Essential JavaScript

The design of SlimWeb is inspired by the observation that not *all* JS might be essential. We selected unicef.org as an example to show the performance gain of removing non-critical JS, without sacrificing the main content or the interactive functionality of the page. SlimWeb indicates that the page embeds 3 analytics, 4 social, and 1 advertising element. We compared the performance of three versions of the unicef.org page:

Table 1: Performance of 3 versions of unicef.org.

| Metric | Original | JS-Free | Optimized |
|---|---|---|---|
| Number of JS requests | 23 | 0 | 5 |
| Size of resources (MB) | 3.8 | 1.2 | 2.0 |
| Speed Index (s) | 4.8 | 2.2 | 3.4 |
| Time to Interactive (s) | 15 | 2.3 | 6.1 |

- *Original*: which is the original version of the page.
- *JS-Free*: where the page was loaded in a browser after disabling JS.
- *JS-Optimized*: where page was loaded by SlimWeb with analytics, advertising and social elements blocked, while other JS elements are preserved.

Table 1 shows the performance results with four metrics. The number of JS requests and the size of resources show the difference in JavaScript utilization, and its impact on the total resources' download size, respectively. The effect of JavaScript utilization on the page performance is shown by the Speed Index [7] and the Time to Interactive [9]. As the table shows, removing analytics, advertising and social elements reduces the number of requests by around 78% (from 23 requests in the original page to 5 requests in the optimized page). This reduction causes a ~50% decrease in the size of resources transferred. A considerable impact was also observed in the Time to Interactive, which decreased from 15 seconds in the original version to 6.1 seconds in the optimized version. The optimized page was visually evaluated in terms of content completeness and functionality. The main content of the page and the functional components were fully preserved in the optimized page. The impact of JS removal has resulted in the removal of an ad element from the top of the page. This example shows potential of identifying and removing non-critical JS from web pages for the benefit of mobile users without affecting the quality of these pages.

## 2.2 User Preferences

To understand the user preferences in removing non-critical JS, we conducted an IRB-approved survey with 306 mobile users from different countries in developing regions. The survey was advertised through social media groups, encouraging users with slow connectivity to participate. We asked the participants about the quality of their network connection, and the class of smartphone devices they use to access the web. In addition, we asked them to select all non-essential elements they are willing to remove from web pages for faster browsing including Advertising, Analytics, and Social elements. These elements were listed with easy-to-understand definitions.

Participants reported the following network connectivity: poor (13%), satisfactory (36.6%), very good (36.9%), excellent (13.3%), and the following personal mobile class, low-end phone that costs around 100 USD (10%), mid-class (46.4%), and high-end (43.4%). Results show that 99.6% of the participants identified at least one JS category as non-essential and are willing to remove the elements that belong to that category from web pages to achieve better performance. Detailed percentages of users preferring to remove each category are shown per network condition in Table 2, and per phone class in Table 3. To select the default non-critical categories of SlimWeb, we picked the categories that exceed 50% of the users preferring to remove them in all network conditions and phone types, resulting in setting Ads and Analytics as non-critical categories.

Table 2: Percentages of users preferring to remove Ads/Analytics/Social (split per network condition)

|  | Excellent | Very Good | Satisfactory | Poor |
|---|---|---|---|---|
| Ads | 92.6% | 89.3% | 72.7% | 82.5% |
| Analytics | 73% | 64.6% | 50% | 57.5% |
| Social | 53.6% | 46% | 22.7% | 22.5% |

Table 3: Percentages of users preferring to remove Ads/Analytics/Social categories (split per phone class)

|  | High-end | Mid Class | Low-end |
|---|---|---|---|
| Ads | 92.4% | 80.2% | 80.6% |
| Analytics | 70.6% | 50% | 64.5% |
| Social | 45.8% | 33.8% | 35.4% |

## 3 RELATED WORK
## 3.1 Reducing the Cost of JavaScript

To handle the cost of JS, web developers utilize uglifiers [25, 30] to reduce the size of JS files before using them in their pages. These uglifiers remove all unnecessary characters (including white spaces, new lines and comments) from these files without changing functionality. Despite the potential enhancement in JS transmission efficiency due to smaller sized files, no speedups are expected at the processing level since the browser has to interpret the entire JS. In contrast, existing JS blocking extensions [6, 18, 24, 35, 46, 58] can aid the users in reducing the amount of JS transferred to their browsers with potential processing speedups. These blockers rely on updating the blocking lists to prohibit a specific category of JS such as ads or trackers, with a main drawback presented in their inability to predict if an "unknown" JS falls into a target category. This is because the rules in their blocking lists are generated by human annotators [62] which hinders effective maintenance. For example, the most popular list consists of around 60,000 rules [63] and is not generalized to handle unseen examples. A recent solution called Percival [17] has shown promising results in blocking ads using deep learning. It operates in the browser's rendering pipeline to intercept images obtained during page execution so that it can flag potential ads. The evaluation of Percival shows non-negligible performance overhead in both Chromium and Brave [10] browsers on desktops, thus raising some questions on their suitability for (low end) mobile devices.

## 3.2 JavaScript Characterization and Analysis

The state-of-the-art approaches for characterizing and analyzing Web-based JS have been focusing on a particular type of JS, such as tracking [37, 49], and advertising [42, 47]. Web tracking is studied in [52], and analyzed using different types of measurements in [37]. A privacy measurement tool is also proposed to characterize online tracking behaviors. In [49], an ML classifier is proposed to identify tracking JS, while another ML approach is proposed in [47] for detecting advertising and tracking resources. These resources are analyzed in [42], with an evaluation of the effectiveness of ad-blocking blacklists. Another main focus of JS characterization is the insecure web practices [32, 33, 44, 66, 68, 69]. In comparison to these approaches, we propose to utilize supervised learning to classify each JS element in a web page into the appropriate category, such that elements that fall under the critical categories are preserved, whereas non-critical elements are blocked. On the other hand, the practical performance of JS is rarely examined [60]. In [28], a JS reference interpreter is presented to produce execution traces and interactively investigate these traces in a web browser. In [67], an empirical study for the non-deterministic behavior of JS is performed to accelerate page load, by saving a snapshot of the objects during the first load and copying them to the heap in the next loads. Instead of examining the exact behavior of JS, we provide an ML-driven approach to classify JS elements in web pages, such that non-critical elements are blocked while critical elements are preserved.

## 3.3 Web Complexity Solutions

An increasing interest is shown during the last decade to tackle the complexity of web pages including fast browsers, in-browser tools, and platform-specific solutions. Opera Mini [2] and Brave [10] are two of the most popular web browsers that can be installed on mobile phones for faster browsing. With Opera Mini, instead of processing JS on the user's device, the processing takes place on a server, which sends the resulting page to the user's browser. This requires the browser to repeatedly communicate with the server whenever a JS function is called. Since most interactions with the page result in calling JS functions, each of these interactions can lead to a delay if the connection between the server and the browser is poor [61]. As for Brave, predefined blacklists are considered to block ads for faster pages. Such blacklist-based solutions are unable to detect previously-unseen advertising JS elements.

With SpeedReader [40] tool added to Brave, pages suitable for the reader mode can be significantly enhanced, but SpeedReader is not yet available for mobile phones, and does not consider pages that utilize JS for content generation. Wprof [64] attracts the attention to JS a key bottleneck in the page load due to its role in blocking HTML parsing. To speedup PLT, Shandian [65] restructures the page loading process, whereas Polaris [56] detects additional edges for more accurate fetch schedules. Unlike SlimWeb, these solutions require the browser to download, process and execute the entire JS brought by a given page.

In Flywheel [19], a proxy service is proposed to extend the life of mobile data plans by compressing responses between the servers and the browsers. However, improving the page load time through Flywheel is not always possible. BrowseLite [50] is another recent tool that aims to achieve data savings by applying different image compression techniques. However, it assumes the existence of server-side tools to perform the desired compression. Other platform-specific solutions have been proposed to improve the browsing experience, such as Apple News [21], and Instant Articles [38] for Facebook users.

Google has also tackled the complexity of web pages through Accelerated Mobile Pages (AMP) [41], with the impact characterized in [48]. A major difference between SlimWeb and AMP is that the latter provides a framework for developers to create new faster pages, whereas SlimWeb aims to offer lighter versions of pages on-the-fly at the client-end. Because AMP pages do not usually refer to their original versions, it was not possible to compare the performance of SlimWeb to AMP. In a recent work [27], JSCleaner offers a set of simplified web pages via a proxy server, where critical and non-critical JS elements are identified by a rule-based classifier. A comparison to JSCleaner is performed since both the original and the simplified versions of web pages are accessible via the employed proxy.

## 4 SLIMWEB DESIGN

SlimWeb is a solution to speed up page loads and improve the mobile browsing experience. This is achieved via two main components: a JavaScript classification service and a browser plugin (see Figure 1). The classification service uses a supervised learning classifier to categorize JS elements into one of the following categories: advertising, analytic, social, video, customer success, utility, hosting, and content. These categories are commonly used and defined by the experts of the web community [22]. The first three categories are the potential *non-critical* categories, while the rest are *critical* to the page content or interactivity.

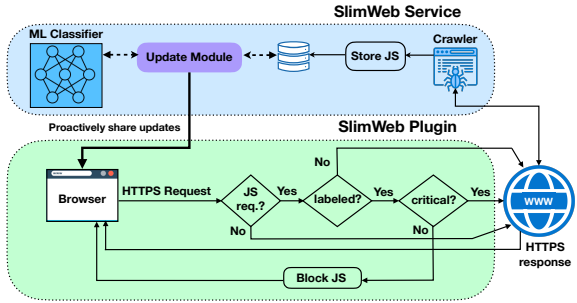

Figure 1: SlimWeb Architecture

SlimWeb's JS classification service crawls popular web pages to identify JS elements used in these pages, and then employs the classifier to label these elements and store their categories in a database. It periodically updates and shares the labels with the users' browser plugins. On the other hand, SlimWeb's browser plugin is responsible for blocking non-critical JS elements. These elements are identified based on the labels received from the service. When a web page is requested by the user, the plugin first checks if a label is locally available for each JS element, such that non-critical elements are immediately blocked. In the case of a label absence, the plugin considers the corresponding JS element as critical and requests it from the web.

## 4.1 JavaScript Classification

SlimWeb employs a novel ML classifier to categorize JS elements embedded in a web page, and then label each of these elements as critical or non-critical according to the user preference. To overcome the challenging task of understanding the non-deterministic behavior of JS [67], the main design goal of SlimWeb is to give an insight into each JS element embedded in a web page, without evaluating the exact behavior of these elements, nor executing its code. The classifier categorizes a given JS element into one of the known JS categories using supervised neural network model, which shows a better and more robust performance over alternatives.

Algorithm 1 shows an overview of the ML classifier. For a given JS element, the classifier outputs a category with a certain accuracy. If the accuracy value is above a predetermined threshold, the algorithm returns the predicted category. Otherwise, it returns *unassigned* category. The returned category is used by the plugin to determine if a given JS element is critical or non-critical according to the user preference. For example, if the classifier returns a *social* category for a given JS element, while the user setting identifies *social* as non-critical, then the element will be considered non-critical. JS

---

**Algorithm 1** JavaScript Classification Algorithm

1: **INPUT** threshold, string script[]
2: **OUTPUT** category
3: **procedure** ASSIGNCATEGORY(*threshold*,*script*)
4:    *category*, *accuracy* = *MLClass(script)*
5:    **if** *accuracy* > *threshold* **then**
6:       **return** *category*
7:    **else**
8:       **return** *unassigned*

---

elements with *unassigned* category are conservatively labeled as critical.

*4.1.1 Models Training.* To select the right ML classifier in SlimWeb, we train an eight class prediction model, which considers eight categories mentioned in [22]. We evaluated the performance of six supervised learning models for JS classification: three distance-based classifiers (K-Nearest Neighbors (KNN), Support Vector Machine (SVM) and Linear Support Vector Classifier (LSVC)), two tree-based classifiers (Random Forest Classifier (RFC) and XG Boost) [29], and a simple neural network model. Evaluations are performed on a desktop machine with Intel core i7 CPU @ 3.60GHz x8, 16GB RAM. To evaluate the performance of the simple 4-layer neural network model, we build a network with an input layer, two hidden layers and one output layer. To determine the optimal number of hidden layers, we perform hyper-parameter search optimization, where no significant improvement is observed in the accuracy when using more than two hidden layers. The *ReLU* (Rectified Linear Unit) [20] activation function was used in the hidden layers while the output layers utilized the *softmax* activation function. The total number of neurons in the input layer was equal to the total number of features. Based on hyper-parameter search, the optimal number of neurons is set to 350 in the first hidden layer, and 50 in the second hidden layer. The learning models are compared using *Recall*, *Precision* and *F1-Score* as performance metrics.

*4.1.2 Model Selection.* The detailed evaluation of the models across the different categories is shown in Figure 2. The results of the SVM model highlighted by Figure 2b shows the lowest recall of 63% in predicting the social category, while the other categories have a recall higher than 79%. Figure 2a presents the results of the KNN model, where a much lower performance is shown in predicting the social category. In contrast, LSVC (Figure 2c) shows a slightly better performance compared to KNN. The results of both of the tree-based classification models considered in this evaluation, which are RFC and XG Boost, shown in Figures 2d and 2e, respectively, show similar performance in terms of precision, recall, and f1-score.

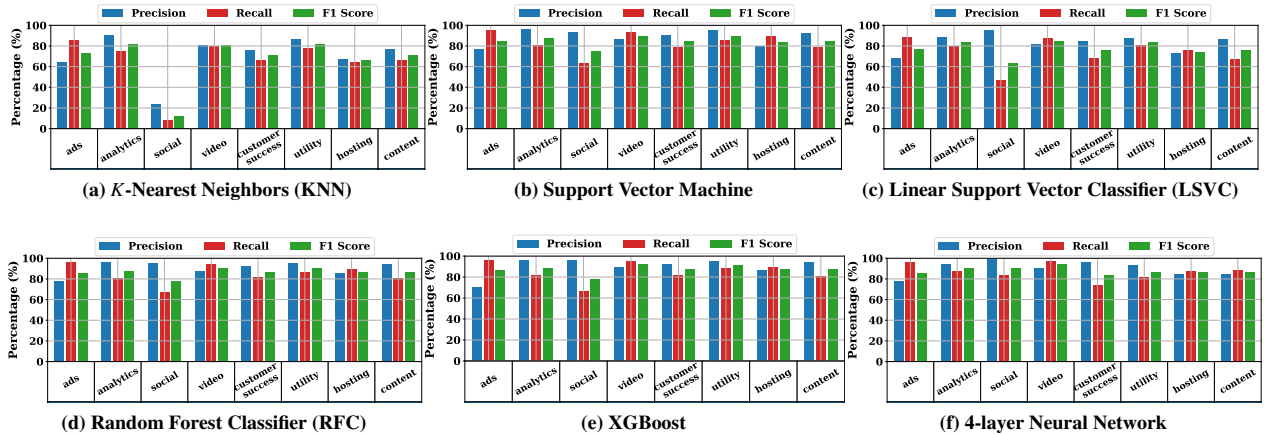

Figure 2: ML classification Models overall comparison

Table 4: A summary of ML models evaluation

| Metrics | KNN | SVM | LSVC | RFC | XG-Boost | NN |
|---|---|---|---|---|---|---|
| Recall | 0.75 | 0.86 | 0.79 | 0.88 | 0.88 | 0.89 |
| Precision | 0.76 | 0.88 | 0.81 | 0.89 | 0.89 | 0.9 |
| F1-Score | 0.75 | 0.86 | 0.79 | 0.88 | 0.88 | 0.89 |

Table 4 presents a summary of the evaluation results of the six models. Each value in the table represents a weighted-average of the corresponding metric, which is computed across all the 8 different categories. In general, tree-based classification models provide better performance in comparison to the distance-based models (KNN, SVM, and LSVC), while the neural network model achieves the best performance (with a slight improvement in precision and recall compared to RFC and XG Boost). Consequently, we select it for JS classifier in SlimWeb. For a multi-class classification problem with 8 classes, a 90% precision-recall achieved by a simple neural network is quite significant; we attribute this accuracy due to the size of the dataset of 127,000 JS elements and believe that this can improve with a larger dataset.

### 4.2 Dataset

We created a dataset to train the supervised learning models, by crawling 20,000 popular pages from [1] and caching JS elements used by these pages. For each JS element, we extract the domain that corresponds to it for a potential match with an entity in a recent HTTP Archive repository [45], which provides a set of entities representing existing JS libraries with their associated domains and categories. These categories are identified and commonly used by experts in the web community [22]. Our dataset consists of 127,000 JS element that were labeled with categories out of 500,000 JS elements crawled from 20,000 popular web pages. The category of a given JS element is used to determine the criticality of that element and decide to block or not to block it accordingly. Each JS element is represented as a vector of features according to [27] with a category label. These features cover the comprehensive set of APIs [53, 54] to interact with the Document Object Model (which defines the structure of HTML documents and the way in which they can be accessed and manipulated). While other representations can take the structure of the code into account, it is shown that the text representation is sufficient [59]. To learn effectively over a small training dataset, we apply Recursive Feature Elimination (RFE), which is a feature selection method that removes the less informative features, resulting in a list of 508 features instead of 1262.

The question here might be: why not matching the URLs of the JS elements in web pages to know their categories instead of utilizing an ML classifier. The answer is that the model would serve the objective of predicting the categories of unknown JS elements embedded in web pages. Moreover, developers might select to locally host some of the known JS, where the local URLs would not match with the elements in the repository. Although the created dataset is limited to well-known JS libraries, it serves a large portion of JS in today's web due to the following reasons: 1) it combines the features (not the URLs) of the elements with their associated categories. 2) The utilized features are associated with the JS access to web pages, which makes them appropriate for category prediction of previously unseen web-based JS.

## 5 IMPLEMENTATION & DEPLOYMENT

On the user's browser, SlimWeb uses a custom plugin that assigns a class to each JS element (either critical or non-critical) according to the user preference. Motivated by the results of the user preference survey presented in Section 2.2, the plugin considers ads and analytics as non-critical categories by default, while other categories are considered critical. Users are allowed to alter the default settings according to their preferences, and to have different configurations on a per-page basis according to their browsing experience. The plugin's local storage is implemented using IndexedDB API to store the labels of the JS resources with a capacity limited to 50 MB. The database consists of a sequence of JS entries, where each entry is associated with a URL and an assigned label. The SlimWeb plugin is implemented using JavaScript for Firefox browser, and is accepted by Mozilla as a Firefox extension. By default, SlimWeb is deployed as a server-based solution as described in Section 4, where JS elements are classified by a server-side service that regularly updates and shares the classification results with the users' browsers.

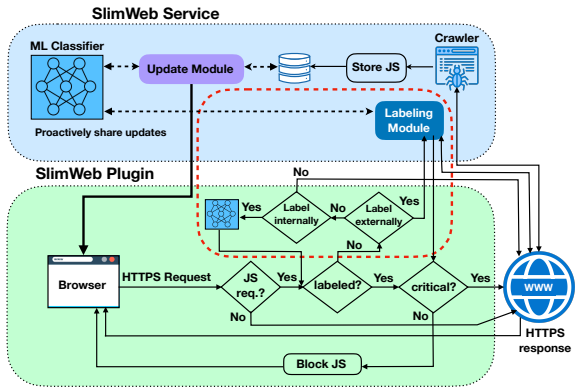

**Figure 3: Alternative SlimWeb deployment options**

Figure 3 shows alternative deployment strategies (inside the dotted red area). The first strategy is to allow the plugin to contact SlimWeb's server in the case of JS labels *misses* in the plugin cache (that is, elements whose labels are not part of the updates shared by the server). Such requests can leak potentially private user information, since it allows the server to identify the actual web page being accessed by the user. However, we argue that this privacy concern is not worrisome for the following reasons. First, SlimWeb asks for the user permission before attempting to classify an unlabeled element at the server. Second, the case of missing labels is rare, since SlimWeb regularly propagates labels of JS elements used in popular pages to avoid extra delays (proportional to the latency between the plugin and the server plus the time needed by the server to fetch and analyze such JS), and to reduce the privacy exposure. Third, JS are commonly shared between pages, e.g. jQuery is used by 73% of the 10 million most popular websites, and websites are encouraged not to host local versions but rely on CDN-hosted ones. The combination of limited JS uniqueness with SlimWeb's goal to achieve a high cache hit rate for JS labels makes us confident of SlimWeb's limited privacy invasion for this deployment.

A third deployment option is to integrate the classification engine directly into the browser. Indeed, some of our early testing have demonstrated that it is possible to run the JS classification on a mobile phone browser using TensorFlow [16]. However, such an approach comes with its own limitations. For example, since the classification engine has to operate at the client side, the content of a given JS file has to be downloaded first by the browser, thus defying the data saving goals for unknown JS. Nevertheless, once the JS element is classified, the class label will be internally stored within the plugin cache to be used later. This means that a JS element would only need to be classified once, and the label will be known in subsequent requests to the same JS.

## 6 EVALUATIONS

This section evaluates SlimWeb with respect to objective metrics like how fast it can load webpages, data savings, and web compatibility. We use two low-end mobile devices: a HUAWEI Y511-U30 (which costs about $89, and is equipped with a dual-core 1.3 GHz Cortex-A7 CPU and 512 MB RAM) and a Xiaomi Remdi Go (which costs about $70, and is equipped with a quad-core 1.4 GHz Cortex-A53 CPU and 1 GB RAM). The Xiaomi phone connects to the Internet over a fast WiFi where network throttling was used to emulate both 3G (downlink/uplink set to 1.6 Mbps/768 Kbps, RTT set to 150ms) and 4G (downlink/uplink set to 12 Mbps, RTT set to 70ms) networks, whereas the HUAWEI phone connects to the Internet using real 3G/4G networks. The two phones are located in two different labs, whose locations were anonymized to respect the double-blind review policy. Each mobile device connects via USB to a Linux machine which is used to run browser automation tools: Browsertime [51] with the Firefox browser and Lighthouse [11] with both Chrome and Brave. Such browser automation tools are used to automate both web page loads and telemetry collection, e.g. performance metrics and network requests. Our testbed further consists of a second Linux machine which acts as SlimWeb's server and runs the JS labeling service (see Figure 1). This machine was located in the same lab as the Xiaomi Remdi Go and about ∼100 ms (median latency) away from the HUAWEI Y511-U30.

Due to the lack of addon support for Chrome and Brave in Android, to evaluate SlimWeb with these browsers we opted for a proxy solution. We implemented SlimWeb client-side capabilities in the mitmproxy [12] by allowing it to operate on encrypted traffic installing a root certificate authority[1] on the device under test. It follows that the testbed consists of another Linux machine acting as a proxy which is located in the same LAN to minimize latency. This machine is employed only to extend our experiments to Chrome and Brave, and is not a core component of SlimWeb. All our evaluations were performed over different subsets from the 11,000 unique pages collected in [23], which consists of the anonymized Chrome browsing history of 82 undergraduate students from Pakistan in a period of 100 days. These are popular pages from a developing region, where low-end devices are common.

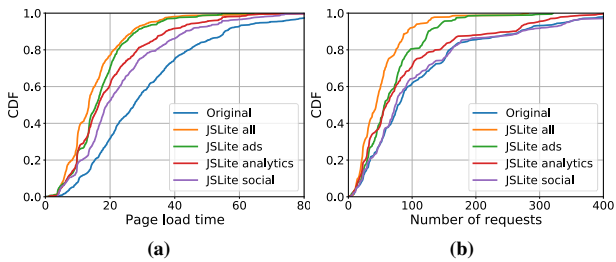

**Figure 4: SlimWeb configurations: Quantitative results**

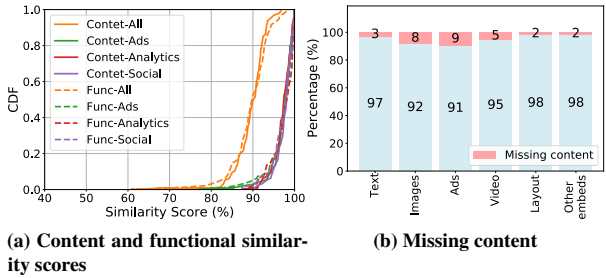

(a) Content and functional similarity scores

(b) Missing content

**Figure 5: SlimWeb configurations: Qualitative results**

## 6.1 SlimWeb Configurations Evaluation

The user preference survey indicated that users are willing to block the following JS categories: ads, analytics, and social. To understand the effect of blocking these different JS categories we devised three evaluations, blocking a different JS category each time, in addition to a final evaluation that blocks all of the three categories. In this experiment, we automate the Chrome browser using Lighthouse to load the 95 most popular pages from the 11,000 unique pages collected in [23]. We have used the Xiaomi phone over an emulated 3G network and with an all requests with the SlimWeb proxy – due to the lack of plugin support in Chrome. Each experiment was repeated 10 times.

[1] https://docs.mitmproxy.org/stable/concepts-certificates/

Figure 4 shows the Cumulative Distribution Functions (CDFs) for page load time, and number of network requests. Figure 4a shows that blocking ads only (green curve) achieves the highest PLT reductions compared to the other two JS categories, which is almost comparable to blocking all JS categories (orange curve). The figure also shows that social has the least impact on the PLT compared to the other categories, followed by blocking analytics. Although blocking ads achieves the closest reductions to blocking all categories, it does not come close in terms of reducing the number of network requests (Figure 4b). This suggests that it is beneficial for users in developing regions who care about their data consumption to block all three categories to achieve the most data savings, with more than 50% reduction in page size at the median (from 1300 KB to 600 KB). Additionally, reducing the number of network requests indirectly improves the phone energy consumption, given that it reduces the usage of the cellular interface.

In order to evaluate the effect of SlimWeb on the quality of the pages, we conducted a user study with 62 participants. These participants were recruited from an international university, and each was given a random set of 10 pages out of the full set, and was asked to compare the four different versions of the pages to their original counter part in terms of content similarity and functionality. The users gave scores out of 10. An institutional review board (IRB) approval was given to conduct the user study, and all the team members have completed the required research ethics and compliance training, and were CITI [5] certified.

Figure 5a shows the CDFs of the content similarity (represented by the solid lines) and the functional similarity (represented by the dashed lines) to the original pages in a form of a percentage score for each SlimWeb configuration. It can be seen that all individual blocking configurations achieve similar scores of more than 90% similarity for both content and functionality. While a combination of all the configurations shows a lower similarity score, the overall scores are still above 90% for half of the pages, and the rest of the pages achieve more than 85% similarity. Finally, we show the percentage of pages missing content split across the different content types in Figure 5b. The figure shows that users have reported that about 9% of the pages are missing ads, followed by 8% of pages with missing images. The rest of the categories are only missing in a very small percentage of pages $< 5\%$.

## 6.2 Comparison to JSCleaner, Privacy Badger, and ADBlock

Here, we compare SlimWeb with several key "competitors": JSCleaner, and Privacy Badger combined with ADBlock. JSCleaner [27] is a recent solution that offer simplified pages via a proxy server (see Section 3). Privacy Badger [39] is a browser extension that blocks both advertising and tracking, whereas ADBlock [18] is one of the most popular ad-blocking browser extension that is used by tens of millions of users worldwide. Our comparison is both *quantitative*, i.e., focusing on speed and data usage, and *qualitative*, i.e., focusing on similarity of the lightened pages with respect to the original pages. We leverage Firefox browser using the HUAWEI phone equipped with the SlimWeb plugin. Browsertime is used for automation and no proxy is used given that Firefox mobile supports addons – with the exception of JSCleaner which is a proxy-based solution. We consider real 3G and 4G network conditions to evaluate the top 500 most popular pages from [23].

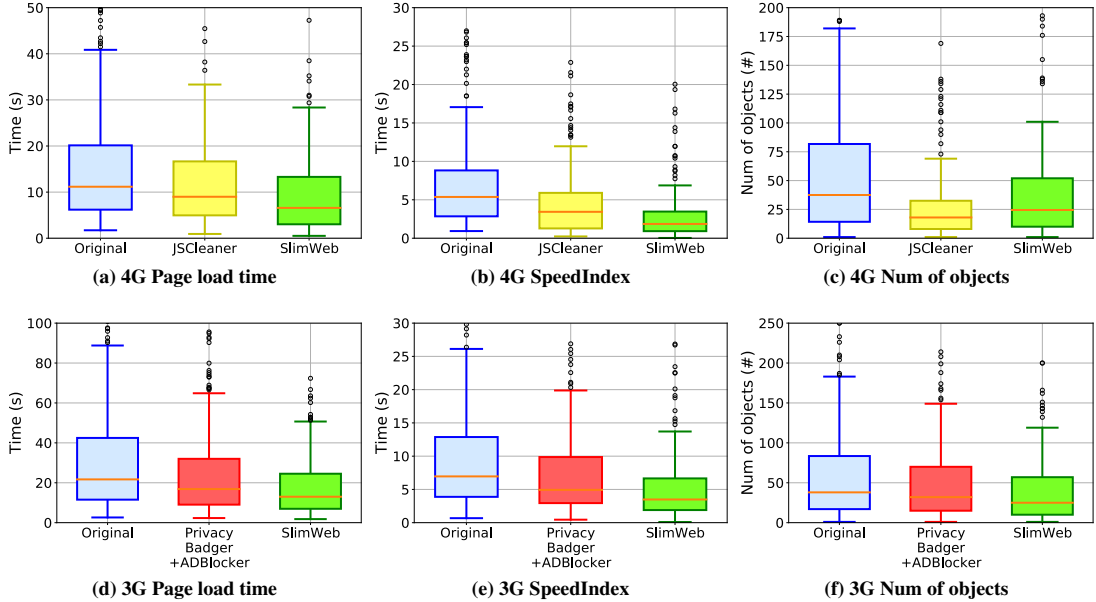

Figure 6: SlimWeb quantitative results for 3G and 4G

*6.2.1 Quantitative Comparison.* The following metrics are considered in the quantitative evaluation: a) Page Load Time, b) Number of objects per page, and c) Speed Index (which is an important user experience metric). Figures 6a to 6c show the performance of SlimWeb in comparison to both: the original and the JSCleaner pages over the 4G cellular network. Figure 6a shows that SlimWeb significantly improves the PLT in comparison to the original and the JSCleaner pages. Specifically, SlimWeb reduces the PLT by more than 50% in comparison to the original, and by around 30% in comparison to JSCleaner. This can observed by comparing the median value of SlimWeb (green box) to the original (blue box) and the JSCleaner (yellow box). Figures 6b and 6c show that SlimWeb reduces the SpeedIndex by more than 60%, and the number of web objects by more than 40% in comparison to the original. Despite that SlimWeb preserves more objects in comparison to JSCleaner, it shows an improvement in the user experience by an additional 30% reduction in the SpeedIndex.

Figures 6d to 6f show the performance of SlimWeb over 3G network in comparison to the original and the privacy-badger+AdBlock. Results show about 50% reduction in the PLT when using SlimWeb (green box), in comparison to a reduction of 25% in the privacy-badger+AdBlock (red box). Similar observations can be seen in Figures 6e and 6f, where SlimWeb is proven to achieve a significant reduction in the number of objects and SpeedIndex in comparison to the original and the privacy-badger+AdBlock. In conclusion, results presented in Figure 6d to 6f prove that the improvement seen in SlimWeb cannot be achieved by existing blocking solutions, even when using a combination of these solutions together. This is due to the use of supervised learning in SlimWeb to classify JS elements, while other solutions rely on blocking lists that ignore previously-unseen elements.

*6.2.2 Qualitative Comparison.* For qualitative evaluation, we compute the similarity scores of pages lightened by SlimWeb, JSCleaner, PrivacyBadger+AdBlock with respect to the original pages. The similarity score of each version is computed using *PQual* [43], which is a tool that automates the qualitative evaluation of web pages using computer vision. Figure 7

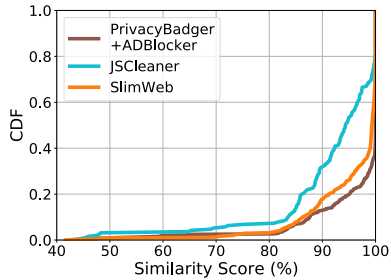
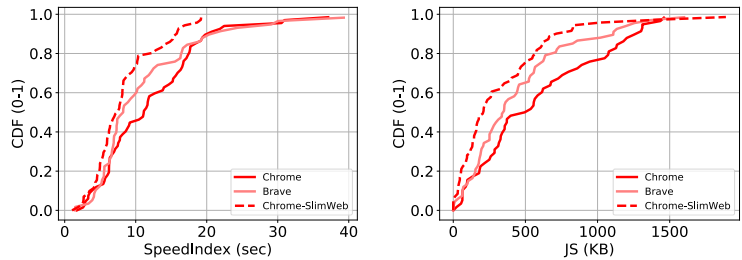

Figure 7: SlimWeb qualitative evaluations (PQual similarity)

(a) CDF of SpeedIndex (sec)

(b) CDF of JS sizes (KB)

Figure 8: Comparing SlimWeb with Chrome and Brave

compares the similarity scores of each page versions as CDFs. The figure shows that more than 80% of the SlimWeb pages have a score that exceeds 90%. Results confirm that SlimWeb does not affect the overall content and the functionality of the pages by blocking the non-critical JS. In contrast, the simplification process of JSCleaner is affecting the quality of a higher percentage of pages. Results also confirm that there is only a small negligible difference in the similarity scores between the SlimWeb and PrivacyBadger+AdBlock.

### 6.3 Comparison With Chrome and Brave

In this section, we investigate SlimWeb's performance using the Chrome (version 87.0.4280.101) and Brave (version 1.18.77) browsers, respectively the most popular Android browser [55] and an upcoming privacy-preserving browser with integrated ad-blocking capabilities [10]. Due to the lack of plugin support for both Chrome and Brave in Android, we resort to SlimWeb's proxy for these experiments. Browser automation is realized via Lighthouse and a 4G network is emulated. The same corpus of pages used in Section 6.1 is also used here, and the test device is the Xiaomi phone.

Figure 8 summarizes our results with respect to performance – due to space limitations we only report on SpeedIndex but similar observations hold for other metrics – and JS usage. We report results for both regular Chrome and Brave, i.e., when the browsers are directly connected to the Internet, and for *Chrome-SlimWeb* where Chrome's traffic is intercepted by SlimWeb's proxy. We also experimented with Brave via our SlimWeb proxy; results are omitted since we observed artifacts from the interaction between Brave's adblock and our proxy. Figure 8a shows significant acceleration offered by Chrome-SlimWeb compared to both regular Chrome and Brave. For example, the median SpeedIndex reduces by 4 and 1 seconds when comparing Chrome-SlimWeb with Chrome and Brave, respectively. The lower reduction observed when comparing with Brave is due to its native adblocker which already reduces web page complexity offering significant speedups compared to Chrome [4].

Figure 8b verifies the latter observation by showing the CDFs of the bytes of JS files transferred during these experiments. The figure shows that Brave serves overall lighter pages compared with Chrome, e.g., median JS size reduces from 446 to 337KB ($\sim$25% reduction). However, SlimWeb manages to offer additional 25% savings, e.g., reducing the median JS size down to only 204KB. These benefits come at the expense of marginal loss of content/functional similarity to the original pages (10-15% reduction shown earlier in Figure 5a). In Section 6.4.2, we demonstrate that many users are willing to use SlimWeb even with the marginal loss in similarity. By downloading only critical JS, SlimWeb achieves better PLT in comparison to both Chrome and Brave.

### 6.4 Evaluation with Real Users

Real users play a fundamental role in SlimWeb's success. Therefore, it is important to evaluate it with real users and assess their willingness to help correct potential web compatibility issues caused by SlimWeb. Our evaluation with real users include a user study encompassing 20 participants who installed and run SlimWeb's plugin on Firefox over 14 days, and a survey covering 588 volunteers.

*6.4.1 SlimWeb User Study.* We evaluated SlimWeb by requesting 20 users to install the plugin on their smartphones and browse the web from these devices for a period of two weeks. Participants were then requested to respond to a set of survey questions by the end of the evaluation period. Out of 20 participants, 19 respond to the survey questions. All participants believe that SlimWeb is useful in providing lighter versions of web pages. Results show that the plugin helps 63% of the users to browse more pages with their data plan. Around 68% of the users agree that the feedback on broken pages helps in improving the performance of the plugin. The users were asked to rate the following on a 0-10 scale [26]: the usefulness of the plugin, the content completeness, the functional and interactivity features, the performance and the quality of the lightened pages. Figure 11 depicts a summary of the users' responses in a percentage score, where a median of at least 80% for all of the requested assessments is shown.

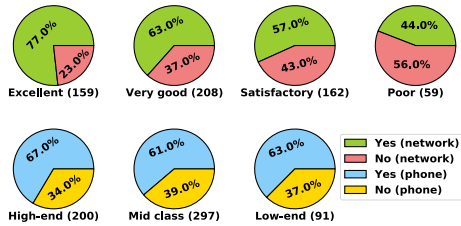

Figure 9: Willingness to use SlimWeb, split as a function of network type (top), and phone class (bottom).

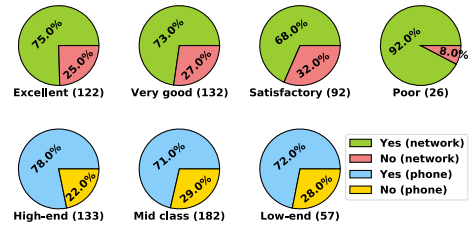

Figure 10: Willingness to fix broken pages, split as a function of network (top), and phone class (bottom).

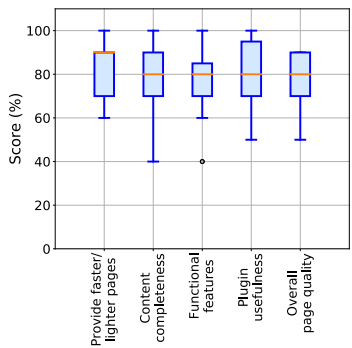

Figure 11: SlimWeb user evaluation scores

*6.4.2 SlimWeb Usage Willingness Survey.* A survey is conducted to understand the willingness of larger number of users to utilize SlimWeb. Users are asked to indicate their connectivity condition and phone type. Further, they are introduced to the concept of the SlimWeb's plugin as a solution to speed up web page loads and improve their browsing experience, which might come at the expense of removing certain components from the pages, and/or breaking some of them. After that, the users are asked to answer two questions. The first question asks if the user is willing to use the plugin assuming that it might break around 15 to 20% of the pages (some pages might be missing certain components, and in some off-cases the pages might be broken). Two examples of broken pages are shown to clarify how broken pages might look like (in each example, the original page is shown side by side with the lightened broken page). The second question asks if the user is willing to put in the effort to fix a broken page (assuming that the plugin gives the option to bring some of the elements back to the page, by listing all of the blocked elements via an interface and allowing to disable the blocking of each of them to test the page afterwards).

We first utilized google surveys [14] as a paid platform to recruit users across 4 different developing countries. Additionally, we posted the survey on multiple social media groups to reach out to users in countries that are not available for selection in google surveys. We offered the survey in three languages (Arabic, English and Spanish). A total of 588 participants responded to the survey from multiple developing countries including: Philippines, Ghana, South Africa, Pakistan, Colombia, Kenya, Nigeria, Egypt, Yemen, Syria, and Iraq. Responses to the first and the second question are summarized in Figures 9 and 10, respectively (split per network type in the first row of each of the figures, and per phone class in the second row). As Figure 9 shows, most users are willing to use SlimWeb despite the probability of breaking pages, were the highest percentages are shown with excellent connectivity and high-end phones. Most users are also willing to put in the effort to fix the broken pages as Figure 10 shows, where poorly connected users present the highest percentage.

## 7 CONCLUSION

This paper proposes and evaluates SlimWeb, a novel ML-driven approach that automatically derives lightweight versions of web pages by eliminating non-critical JavaScript. SlimWeb aims to improve the web browsing experience of less fortunate users by predicting the class of each JS element, such that critical elements are preserved while non-critical elements are blocked. Evaluation results across a broad spectrum of 500 web pages across 3G and 4G networks show that SlimWeb outperforms existing solutions while maintaining a high similarity to the original web pages. A user study with a group of 20 users shows the practical utility of SlimWeb for mobile users in developing regions. In future work, we aim to perform a large-scale roll-out of SlimWeb.